\documentclass[aps, pra,reprint,twocolumn,english,showpacs]{revtex4-2}
\usepackage{geometry}
\usepackage{ragged2e}
\usepackage{xcolor}
\usepackage{babel}
\usepackage{units}
\usepackage{amsmath}
\usepackage{amssymb}
\usepackage{stmaryrd}
\usepackage{graphicx}
\usepackage{wasysym}
\usepackage{bbold}
\usepackage{mathtools}
\usepackage{blkarray, bigstrut}
\usepackage{float}
\usepackage{makecell}
\usepackage[bookmarks=false]{hyperref}
\newcommand{\DQC}{Duke Quantum Center, Department of Electrical and Computer Engineering and Department of Physics, \\Duke University, Durham, NC 27708}

\newcommand{\ket}[1]{\vert{#1}\rangle}
\newcommand{\Ba}{$^{138}$Ba$^+$} 

\begin{document}
\preprint{APS/123-QED}

\title{Non-invasive mid-circuit measurement and reset on atomic qubits}

\author{Zuo-Yao Chen}
\thanks{These authors contributed equally to this work.}
\affiliation{\DQC}
\author{Isabella Goetting}\thanks{These authors contributed equally to this work.}
\affiliation{\DQC}
\author{George Toh}
\affiliation{\DQC}
\author{Yichao Yu}
\affiliation{\DQC}
\author{Mikhail Shalaev}
\affiliation{\DQC}
\author{Sagnik Saha}
\affiliation{\DQC}
\author{Ashish Kalakuntla}
\affiliation{\DQC}
\author{Harriet Bufan Shi}
\affiliation{\DQC}
\author{Christopher Monroe}
\affiliation{\DQC}
\author{Alexander Kozhanov}
\affiliation{\DQC}
\author{Crystal Noel}
 \email{crystal.noel@duke.edu}
\affiliation{\DQC}
\date{\today}

\begin{abstract}
Mid-circuit measurement and reset of subsets of qubits is a crucial ingredient of quantum error correction and many quantum information applications. Measurement of atomic qubits is accomplished through resonant fluorescence, which typically disturbs neighboring atoms due to photon scattering.
We propose and prototype a new scheme for measurement that provides both spatial and spectral isolation by using tightly-focused individual laser beams and narrow atomic transitions. 
The unique advantage of this scheme is that all operations are applied exclusively to the read-out qubit, with negligible disturbance to the other qubits of the same species and little overhead. 
In this letter, we pave the way for non-invasive and high-fidelity mid-circuit measurement and demonstrate all key building blocks on a single trapped barium ion.
\end{abstract}

\maketitle
\section{Introduction}

Atomic qubits are unsurpassed carriers of quantum information because of their natural indistinguishability, long coherence times \cite{wang2021single}, high-fidelity operations \cite{an2022high, loschnauer2024scalable, bluvstein2022quantum, revB}, and near-perfect measurement efficiency \cite{Sotirova2024}.
Many exciting areas of quantum information such as coherent noise detection \cite{singh2023mid}, quantum many-body simulations \cite{Foss-Feig2021, Schmale2024}, measurement-induced quantum phase transitions \cite{Li2019, Skinner2019, Gullans2020,  noel2022measurement, Koh2023}, and quantum error correction \cite{Shor1995}, require mid-circuit measurement and reset (MCMR) operations that extract information from only a subset of qubits while preserving coherence in other data qubits. MCMR is particularly challenging in atomic systems such as trapped ions and neutral atoms, where even focused laser beams that drive dissipative operations may induce crosstalk with neighboring atoms.

Existing methods for MCMR isolation in atomic systems include the use of multiple atomic species~\cite{negnevitsky2018repeated, singh2023mid}, atom shuttling to well-separated spatial zones~\cite{bluvstein2024mcmr, pino2021demonstration, zhu2023interactive}, and the use of multiple spectrally-distinct qubits in the same atom---the so-called \textit{\textbf{omg}} architecture~\cite{allcockOmgAPL2021,ma2023mcecomg, graham2023mcmomg, Lis2023}. 
The \textit{\textbf{omg}} architecture stands alone as a programmable method for select qubit isolation without the significant overhead of shuttling or particular spatial ordering of multiple species. 

In this work, we propose and demonstrate elements of a novel \textit{\textbf{omg}} method to implement dissipative operations in the middle of a quantum circuit. Tightly-focused laser beams are exploited to spectrally Stark-shift qubit levels in targeted auxiliary atoms to be cooled, measured, or reset. 
While Stark-shifting has previously been used to isolate data qubits from dissipative operations \cite{Norcia2023, hu2025site, Lis2023, yichao}, this involved shifting the data qubits themselves.
In contrast, here the selected and shifted auxiliary atoms are driven through a metastable state and repumped through a strong transition that decays back to the original qubit state, forming a cycling transition for dissipative processes. The other data qubits are well-isolated from these cycles, as the shifts of the auxiliary qubits are set much larger than the (narrow) bandwidth of the transition to the metastable states.
This method of MCMR confers three advantages: (i) all operations are performed with a single atomic species in a stationary arrangement \textit{in situ}, (ii) the data qubits are not shifted or shelved, and (iii) the tightly-focused shifting beams may already be in place for quantum gate operations.

\begin{figure*}[t]
\includegraphics{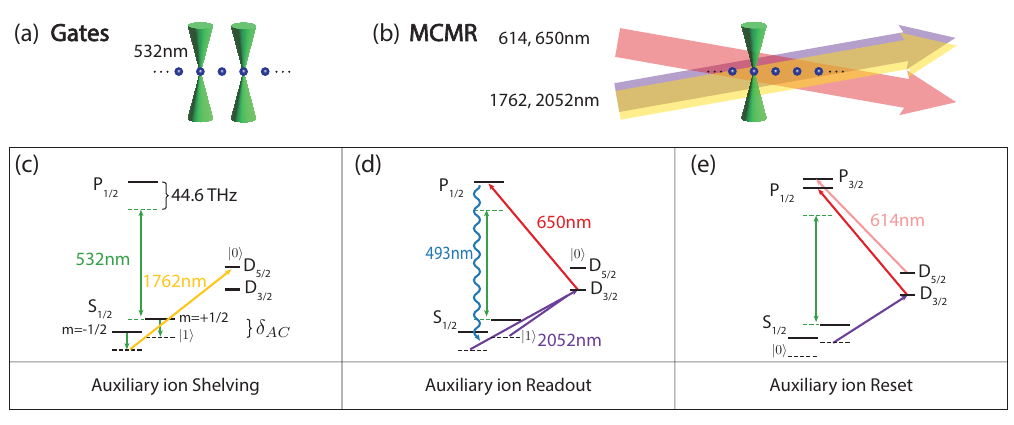}
\caption{\label{fig:mcmrscheme}
Proposed mid-circuit measurement and reset (MCMR) scheme in the atomic \Ba system. (a) Individual-addressing laser beams (at 532 nm for Ba$^+$, for example) conventionally drive stimulated Raman transitions and entangling gate operations \cite{MS, Inlek2017}. (b) Beam layout for executing coherent quantum gates and dissipative mid-circuit operations. MCMR uses the same individual addressing 532 nm beams in (a) but now for Stark shifting selected auxiliary qubits during operations performed by the other global beams. 
(c-e) Sequence of qubit control for mid-circuit operations in \Ba. Ground state qubit levels are Stark shifted by $\delta_{AC}$, with a smaller shift of the $D$ levels (not shown). Selective shelving, measurement, and reset operations are programmed by using the particular beams to shift levels and drive transitions as indicated. 
}
\end{figure*}

\section{Proposed Scheme}

An example of energy level structures supporting this scheme can be found in alkaline earth atomic ions such as Ba$^+$ having $^2$$S_{1/2}$ ground state qubits, $^2$$D_{3/2}$ metastable excited states, and a dissipative cyclic path back to the qubit state through the $^2$$P_{1/2}$ level \cite{lindenfelser2017cooling}.
We specifically consider the $^2$$S_{1/2}$ Zeeman ground state qubit of \Ba, where $|0\rangle \equiv |6S_{1/2}, m=-1/2\rangle$ and $|1\rangle \equiv |6S_{1/2}, m=+1/2\rangle$. It is straightforward to apply this scheme to other qubit types, such as the clock-state qubits in $^{137}$Ba$^{+}$ and $^{133}$Ba$^{+}$. 

Trapped \Ba~qubits can exploit tightly-focused 532~nm laser beams for both conventional quantum gates~\cite{Inlek2017} (Fig.~\ref{fig:mcmrscheme}a) and site-selective Stark shifts at the heart of this MCMR scheme. In order to mitigate intensity crosstalk between adjacent ions, the individual beam waists at the ion positions are set much smaller than the ion pitch~\cite{debnath2016demonstration}. Global beams address the entire chain for cooling, electron shelving and qubit readout as shown in Fig.~\ref{fig:mcmrscheme}b. 
The global beams are tuned to address the shifted transitions to the metastable $D_{3/2, 5/2}$ states, and are therefore off-resonant from the other data qubits in the chain. 

We follow the sequence illustrated in Fig.~\ref{fig:mcmrscheme}c-e:
A 1762 nm laser pulse transfers the population of the qubit state $\ket{0}$ to the long-lived $D_{5/2}$ manifold, making that state dark during fluorescence detection.
To readout the auxiliary, global 2052~nm and 650~nm beams drive the cycling transition from $S_{1/2}$ to $D_{3/2}$ \cite{Kleczewski2012} and through the $P_{1/2}$ manifold, where spontaneously-emitted 493~nm photons are collected \cite{lindenfelser2017cooling}. 
After the mid-circuit measurement, the auxiliary is reset to $\ket{0}$ with 2052~nm and both repump beams (650~nm and 614~nm) (Fig.~\ref{fig:mcmrscheme} e).
\section{Experiment}
To demonstrate the viability of the proposed scheme, we conduct several tests on an existing trapped ion system.
In the experimental setup, we trap a single $^{138}$Ba$^{+}$ ion in a four-rod Paul trap that is not equipped for ion chain operations with individual addressing. 
To implement Stark shifts and to emulate the individual-addressing beams, we focus a single 532~nm beam to approximately 3~$\mu$m x 2~$\mu$m radius and align it to the single ion. 
We first demonstrate Stark-shifted auxiliary readout with a sequence consisting of cooling, shelving, and fluorescence detection. We Doppler-cool the ion for 1~ms and prepare it in $\ket{0}$ with a $10~\mu $s pulse of 493~nm $\sigma^-$-polarized light. To measure the detection fidelity when the ion is in the dark $\ket{0}$ state, we first shelve the ion from $\ket{0}$ to the $|D_{5/2}, m = -1/2 \rangle$ state using a $9~\mu$s $\pi$-pulse of 1762~nm light, followed by 9~ms of fluorescence detection with MCMR light (650~nm and 2052~nm light as shown in~Fig.\ref{fig:mcmrscheme}d). 
To measure the detection fidelity of the bright $|1\rangle$ state, we repeat the procedure but without the 1762~nm shelving pulse, leaving the ion in the $S_{1/2}$ ground state. To address both Zeeman sublevels of $S_{1/2}$, we apply 2052~nm light with two different frequencies to drive both $\Delta m = 0$ transitions from the $\ket{0}$ and $\ket{1}$ states, with $\Omega_{2052}=2\pi\times 100~$kHz. To maximize the power of each frequency tone, we alternate between the two tones, switching every $6~\mu$s. We implement Blackman (BM) pulse shaping \cite{blackman} on the 2052~nm pulses to eliminate frequency sidebands from the fast switching. 

\begin{figure}[h]
\includegraphics[width=0.45\textwidth]{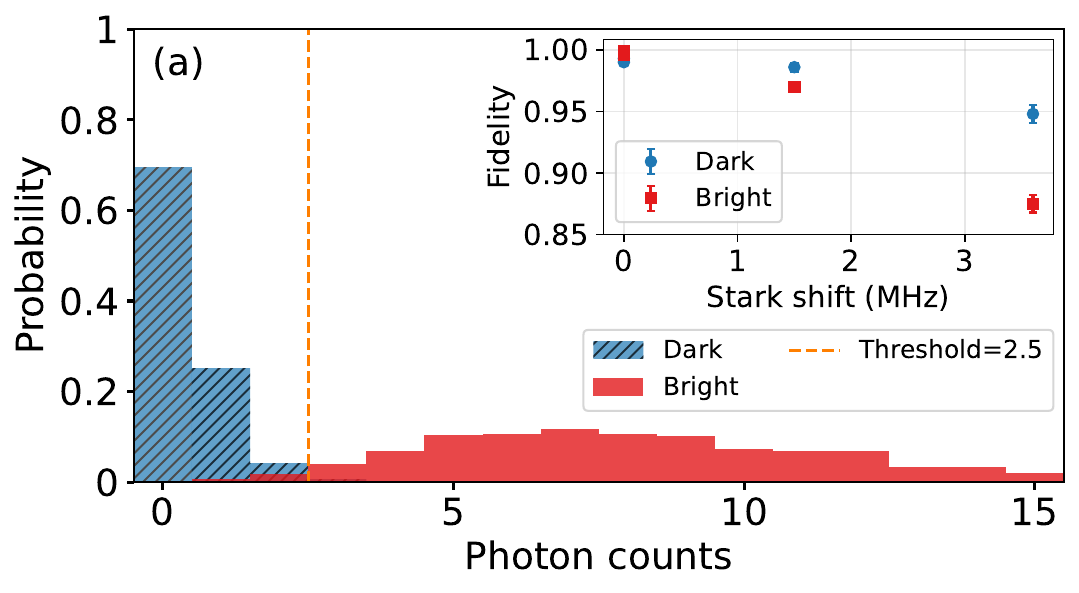} \\
\includegraphics[width=0.45\textwidth]{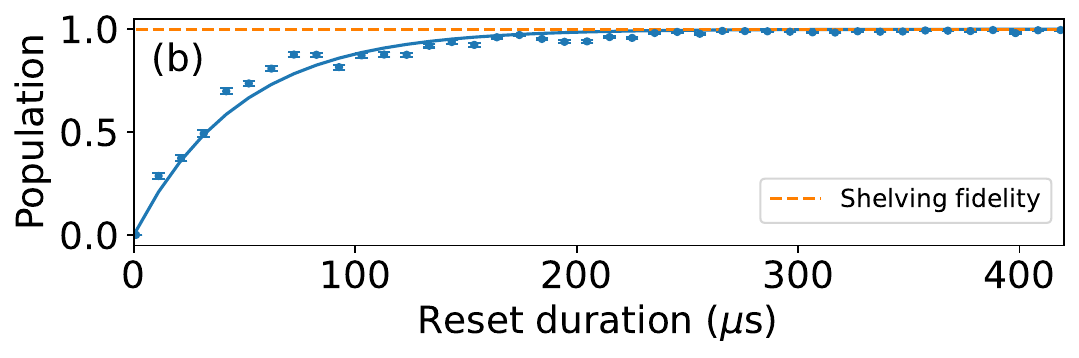} 
\caption{\label{fig:2052_hist} 
Measurement and reset experimental results
(a) Histogram of 493~nm fluorescence counts from detection using MCMR light with a $1.5(2)$~MHz Stark shift. 
Detection fidelities of the dark and bright ion states are $98.6(4) \%$ and $97.0(5) \%$, respectively. 
Inset: As we increase the Stark shift, the ion state detection fidelity decreases due to 532~nm power instability. 
(b) Population in $\ket{0}$ during reset with a 2.6(3)~MHz Stark shift (Fig.\ref{fig:mcmrscheme}e).
The solid blue line is an exponential decay fit with $\tau_r=47~\mu$s. The reset fidelity saturates at the level of fidelity for 1762~nm electron shelving, $99.7(2) \%$ (orange dashed line). 
All error bars are derived using the standard deviation of a binomial distribution, where each point has 1000 shots.}
\end{figure}

Without a Stark shift, we obtain a dark/bright state detection fidelity of $99.0(3)\%$/$99.6(2)\%$ using the un-shifted scheme from Fig. \ref{fig:mcmrscheme}d. With a Stark shift of 1.5(2)~MHz, we measure a dark/bright state detection fidelity of $98.6(4)\%$/$97.0(5) \%$ (Fig.~\ref{fig:2052_hist}a). 
This result is largely limited by intensity fluctuations of the 532~nm beam, which broadens the narrow 2052~nm transition (see Appendix \ref{532stability}).
As the Stark shift is increased by turning up the intensity of the 532~nm light, the bright state detection fidelity degrades from these fluctuations, while the dark state detection fidelity suffers from excess 532~nm scattering background, as shown in the inset of Fig.~\ref{fig:2052_hist}a. 
For standard state-dependent fluorescence detection, we detect for 1 ms with a $99.7(2)\%$ fidelity. This fidelity is limited by our 1762~nm shelving laser.

Next, we measure the reset time and fidelity of the auxiliary qubit when using 2052~nm. 
The reset time is how long it takes to prepare the readout qubit state in $\ket{0}$ after mid-circuit measurement without disturbing the data qubit. Because only one tone of 2052~nm light is applied, no BM pulse shaping is needed. 
To estimate the longest reset time, the ion is pumped to $\ket{1}$ using $10~ \mu$s of 493~nm $\sigma^{+}$-polarized light. Then, we Stark shift the ion by $2.6$(3)~MHz while the 650~nm repump and the $m=+1/2$-addressing tone of 2052~nm are on together for varying amounts of time. We shelve the population in $\ket{0}$ with a 1762~nm pulse and readout the auxiliary using both 650~nm and 493~nm light.
For a sufficiently long reset duration, reset fidelity saturates at the $99.7(2)\%$ fidelity of the 1762~nm electron shelving pulse.
The data is fit to an exponential function, as shown in Fig.~\ref{fig:2052_hist}b, 
and has a characteristic reset time $\tau_r=47 \mu$s.
To reach a reset error of $0.1\%$, a reset time of $326~\mu$s is needed, compared to $127~\mu$s without a Stark shift.
The reset time could be much faster with higher 2052~nm intensity.

\begin{figure}[h]
\includegraphics[width=0.48\textwidth]{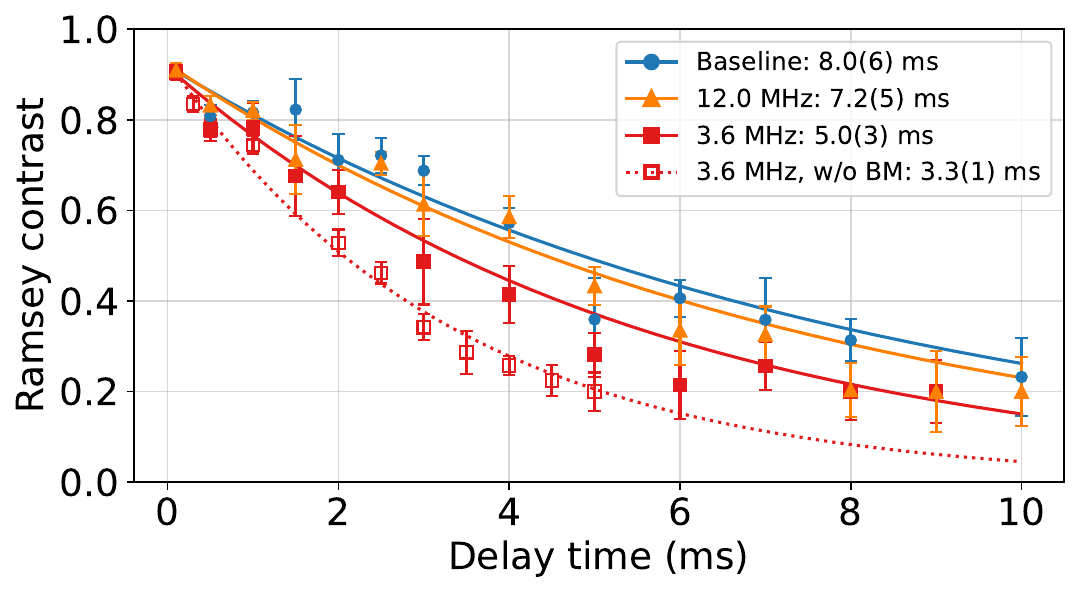}
\caption{\label{fig:coherence_time} 
Time evolution of Ramsey fringe contrast on the 1762~nm transition, indicative of the data qubit coherence during MCMR. The blue circles serve as a baseline without MCMR light. 
The measurement is repeated with MCMR light on during the Ramsey delay time at two detunings of 2052~nm, 3.6~MHz (solid red squares) and 12~MHz (solid orange triangles). The empty red squares show the 3.6~MHz result without Blackman (BM) pulse shaping. The reported coherence times are from experimental fits to the data. With only 12~MHz Stark shift, decoherence is indistinguishable from the baseline. All error bars are derived using the standard deviation of a binomial distribution, where each point has 200 shots.}
\end{figure}

Finally, we conduct Ramsey measurements to determine the coherence time of the data qubit during MCMR. Ideally, the data qubit should maintain perfect coherence and be minimally affected by the measurement and reset sequences. We prepare an optical qubit with 1762~nm light and perform Ramsey spectroscopy with a detuned 2052 beam, mimicking the conditions during MCMR on a Stark-shifted auxiliary ion. 
In a typical system, the individual beam crosstalk on a neighboring data ion is small \cite{debnath2016demonstration, egan2021, HuangCetina2024}, so it is not necessary to turn on the 532~nm beam during the Ramsey free evolution time. 
The results of measuring the data qubit coherence time for different cases are shown in Fig.~\ref{fig:coherence_time}. 

The measurement without 2052~nm exposure serves as a baseline, subject to 1762~nm laser noise and magnetic field drifts, with a coherence time of $8.0(6)$~ms. The 2052~nm detunings are chosen such that the frequency is away from micro-motion sidebands and transitions between the $S_{1/2}$ and $D_{3/2}$ Zeeman sublevels. With a modest detuning of 12 MHz, the coherence time is $7.2(5)$~ms, indistinguishable from the baseline. A smaller detuning of 3.6~MHz shortens the coherence time to $5.0(3)$~ms.
Without the use of BM pulse shaping, the coherence time is reduced to only $3.3(1)$~ms, indicating its efficacy.
These coherence times are limited by fluctuating Stark shifts from the 2052~nm beam, causing dephasing of the qubit during the Ramsey wait time. The fluctuations are dominated by frequency instability of the 2052~nm laser (linewidth of 90(20)~kHz over the bandwidth of the measurement).
With better laser locking to  $<2$~kHz and a detuning of $12$~MHz, we estimate a dephasing time of $>1$~s from this effect.

\begin{figure}[h]
\includegraphics[width=0.48\textwidth]{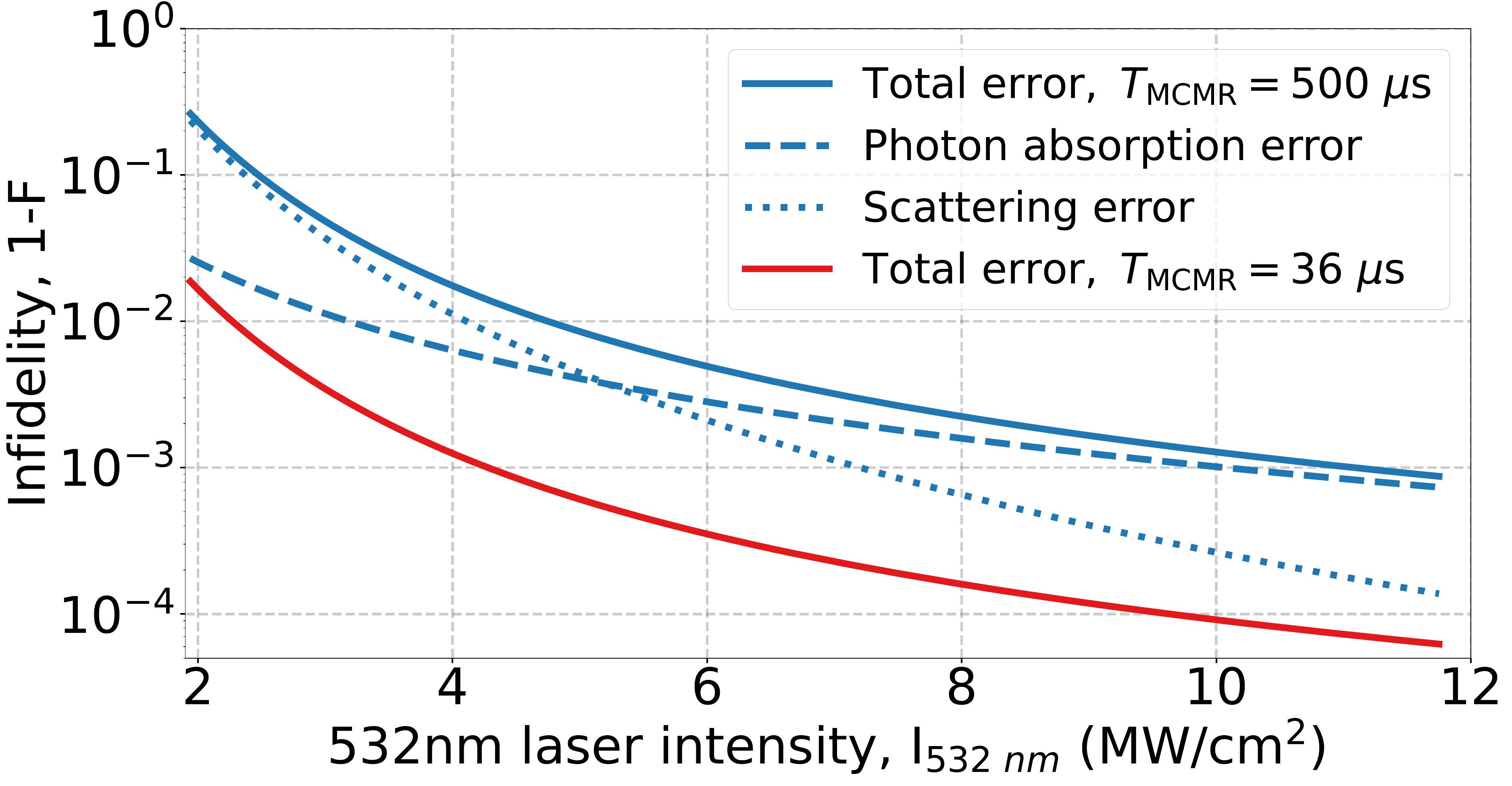}
\caption{\label{fig:fid_vs_SS}Calculated errors for a \Ba~ground state data qubit during MCMR operations as a function of 532~nm laser intensity ($\Omega_{2052}=2\pi\times2~$MHz, $T_\textrm{MCMR}=500~\mu$s). Dashed blue curve is photon absorption from neighboring auxiliary qubit emission at a distance of $d=4$ $\mu$m; dotted blue curve is scattering from global MCMR beams; the solid blue curve is their sum, or the total error. Higher fidelity can be achieved by shortening the detection time with state-of-the-art detection techniques~\cite{crain2019high}, with the total error for $T_\textrm{MCMR}=36~\mu$s shown in the red curve.}
\end{figure}

\section{Practical Considerations}
We now consider practical limits to the fidelity of data qubits under this MCMR scheme. 
The 532~nm addressing beam will give residual Stark-shifts to data qubits similar to that during quantum gate operations \cite{debnath2016demonstration, egan2021, HuangCetina2024}, which will not appreciably change the spectral isolation of data qubits, so we ignore this effect in the following.
Depolarizing errors on the data qubits during MCMR come from two primary sources: (1) photon absorption from 493~nm photons emitted by adjacent ions, (2) scattering from off-resonant excitation into the $S_{1/2}\rightarrow D_{3/2}\rightarrow P_{1/2}$ cycle from the global MCMR beams. Raman scattering on the $S \rightarrow P$ transitions is found to be negligible compared to the above sources \cite{PhysRevA.107.032413}.

First, we consider scattering of the data qubit due to photon absorption from the neighboring auxiliary ion. This rate is given by

\begin{equation}
R = \left(\frac{3\lambda^2}{8\pi^2 d^2}\right)f\beta \Gamma\left(\frac{\beta \Gamma/2}{\delta_{P}-\delta_{S}}\right)^{2}.
\end{equation}
The first term is a geometric factor set by the distance between ions $d$ and the wavelength $\lambda$ of the 493~nm transition. The second term is the scattering rate of the measured ion for an excited state ($P_{1/2}$) fraction $f$. The branching ratio of the $P_{1/2}$ to $S_{1/2}$ decay channel is $\beta=0.73$ and the natural linewidth of the $P_{1/2}$ level is $\Gamma=2\pi\times 20.5$ MHz. We find the optimum excited state fraction is $f=0.04$ by solving a master equation with an assumed $\Omega_{2052}=2\pi\times2$ MHz \cite{marzoli1994laser}. The last term describes the suppression of photon absorption due to the total Stark shift detuning $\delta_{P}-\delta_{S}$ between $S$ and $P$.

Next, we consider the scattering error caused by off-resonant excitation on the cycling transition. The rate of this scattering is determined by $f$, $\Omega_{2052}$, and $\delta_{AC}=\delta_{D}-\delta_{S}$~\cite{marzoli1994laser}.
For a large AC Stark shift $\delta_{AC}\gg\Gamma$, we solve the master equation of the simplified \{$S_{1/2}, D_{3/2}, P_{1/2}$\} three-level system to find the steady-state $P_{1/2}$ population when there is a detuning, and then multiply by $\beta\Gamma$ to find the scattering rate. Note that $\delta_{AC}$ is different among the four Zeeman states because of the tensor shift (see Appendix \ref{app:SS Estimation}); therefore, we use the smallest $\delta_{D}$ for each different 532~nm intensity to provide a conservative estimation on the scattering error.

The total error from photon absorption and scattering depends on the Stark shift applied to the auxiliary ion and the MCMR duration $T_\textrm{MCMR}$.
We show the total error in Fig.~\ref{fig:fid_vs_SS} with $T_\textrm{MCMR}=500~\mu$s, which is comparable to the duration of an entangling gate in a long ion chain \cite{egan2021}. 
Shortening the MCMR duration to $T_\textrm{MCMR}=36~\mu$s, corresponding to a reasonable detection time \cite{crain2019high}(see Appendix \ref{MCMR time}), the total expected error is plotted as the red solid line in Fig. \ref{fig:fid_vs_SS}. 
A Stark shift of more than $150$~MHz has been experimentally demonstrated on \Ba  using a 532~nm laser \cite{lambrecht2017long}, which can be replicated with 96~mW focused down to $2~\mu$m diameter, a suitable size for individual addressing.
Altogether, under these practically achievable conditions, we find a total expected data qubit error below $10^{-3}$.

\section{Conclusion and Outlook}
While we demonstrated all of the building blocks of a non-invasive MCMR scheme on a single ion, this technique is suitable for use with a variety of species of atoms and ions used as qubits in quantum computing systems. Most importantly, this technique can be integrated into a full quantum computing system with minimal overhead when the Stark-shifting laser allowing MCMR is also used to drive quantum gates.

To ensure high-fidelity operations after MCMR, sympathetic cooling (or motional reset) is likely required. Doppler and sideband cooling via the narrow-linewidth quadrupole transition was previously demonstrated \cite{lindenfelser2017cooling}. It is likely possible to use the scheme presented here for site-selective narrow-linewidth cooling since the protocol is similar to fluorescence detection. We leave these experimental explorations to future work with long chains.

\begin{acknowledgments}
The authors are thankful to Boris Blinov for providing the 2052~nm laser and to Jameson O'Reilly for useful discussions. 
This material is based upon work supported by the U.S. Department of Energy, Office of Science, National Quantum Information Science Research Centers, Quantum Systems Accelerator (DE-FOA-0002253). 
Additional support is acknowledged from the DARPA Measurement-based Quantum Information and Transduction program (HR0011-24-9-0357) and the NSF STAQ Program (PHY-1818914). 
A. Kalakuntla is supported by the AFOSR National Defense Science and Engineering Graduate (NDSEG) Fellowship.  
\end{acknowledgments}

\clearpage

\appendix

\newpage

\section{\label{app:SS Estimation}532~nm AC Stark shift estimation}

Considering a laser field with electric field amplitude $E$, frequency $\omega$, and linear polarization $\boldsymbol{\epsilon}$, the operator of AC Stark shift interaction is given by\cite{le2013dynamical}:

\begin{equation}
    H_{AC\ Stark} = -\frac{1}{4}E^{2}[\alpha^{scalar}_{nJ} + \alpha^{tensor}_{nJ} \frac{6(\boldsymbol{\epsilon}\cdot \boldsymbol{J})^{2}-2\boldsymbol{J}^{2}}{2J(2J-1)}]
\end{equation}
where $\alpha^{scalar}_{nJ}, \alpha^{tensor}_{nJ}$ are (dynamic) scalar and tensor polarizability of atomic level in shell $n$ with quantum number $J$. Assuming linear laser polarization, therefore $\boldsymbol{\epsilon}$ being real, the vector shift is zero and already omitted in the above equation. 

The scalar and tensor polarizability is given by:

\begin{equation}
\begin{aligned}
    \alpha^{scalar}_{nJ} &=  \frac{1}{\sqrt{3(2J+1)}}\sum_{n^{\prime}J^{\prime}}(-1)^{J^{\prime}+J+1}\begin{Bmatrix}
        1&0&1\\J&J^{\prime}&J
    \end{Bmatrix} \cdot \\&\frac{6\pi\epsilon_{0} c^{3}(2J^{\prime}+1)A_{J^{\prime}J}}{\omega_{J^{\prime}J}^{2}(\omega_{J^{\prime}J}^{2} - \omega^{2})}
\end{aligned}
\end{equation}

\begin{equation}
\begin{aligned}
    \alpha^{tensor}_{nJ} &= \sqrt{\frac{10J(2J-1)}{3(J+1)(2J+1)(2J+3)}}\sum_{n^{\prime}J^{\prime}} (-1)^{J^{\prime}+J}\cdot\\ &\begin{Bmatrix}
        1&2&1\\J&J^{\prime}&J
    \end{Bmatrix}\frac{6\pi\epsilon_{0} c^{3}(2J^{\prime}+1)A_{J^{\prime}J}}{\omega_{J^{\prime}J}^{2}(\omega_{J^{\prime}J}^{2} - \omega^{2})}
\end{aligned}
\end{equation}
where the curly matrix is Wigner-6j symbol, and the summation runs over all the atomic levels $n^{\prime}J^{\prime}$ that couples to level $nJ$ with decay rate $A_{J^{\prime}J}$ and transition angular frequency $\omega_{J^{\prime}J}$. The results are listed in Table.~\ref{Table: SDP_Polarizability}. We point out that the values we report here are different from \cite{UDportal} because we do not include other high-lying levels (data for transition lines of shorter wavelength are not available). For $S_{1/2}-D_{3/2}$ transition and 493~nm photon ($S_{1/2}-P_{1/2}$ transition) considered in this paper, the overall AC Stark shift mostly comes from the scalar shift of $6S_{1/2}$, so the difference is only by a few percent and does not affect the effectiveness of our scheme.

An external B field is usually applied along the $z$ direction to define quantization axis and lift the Zeeman degeneracy. When the laser field is propagating along B field in $z$ direction, its linear polarization is perpendicular and we define the $x$ direction along the laser polarization. The AC Stark shift operator is then reduced to:

\begin{equation}
    H_{AC\ Stark} = -\frac{1}{4}E^{2}[\alpha^{scalar}_{nJ} + \alpha^{tensor}_{nJ} \frac{6\boldsymbol{J_{x}}^{2}-2\boldsymbol{J}^{2}}{2J(2J-1)}]
\end{equation}

\begin{figure}[b]
\includegraphics[width=0.48\textwidth]{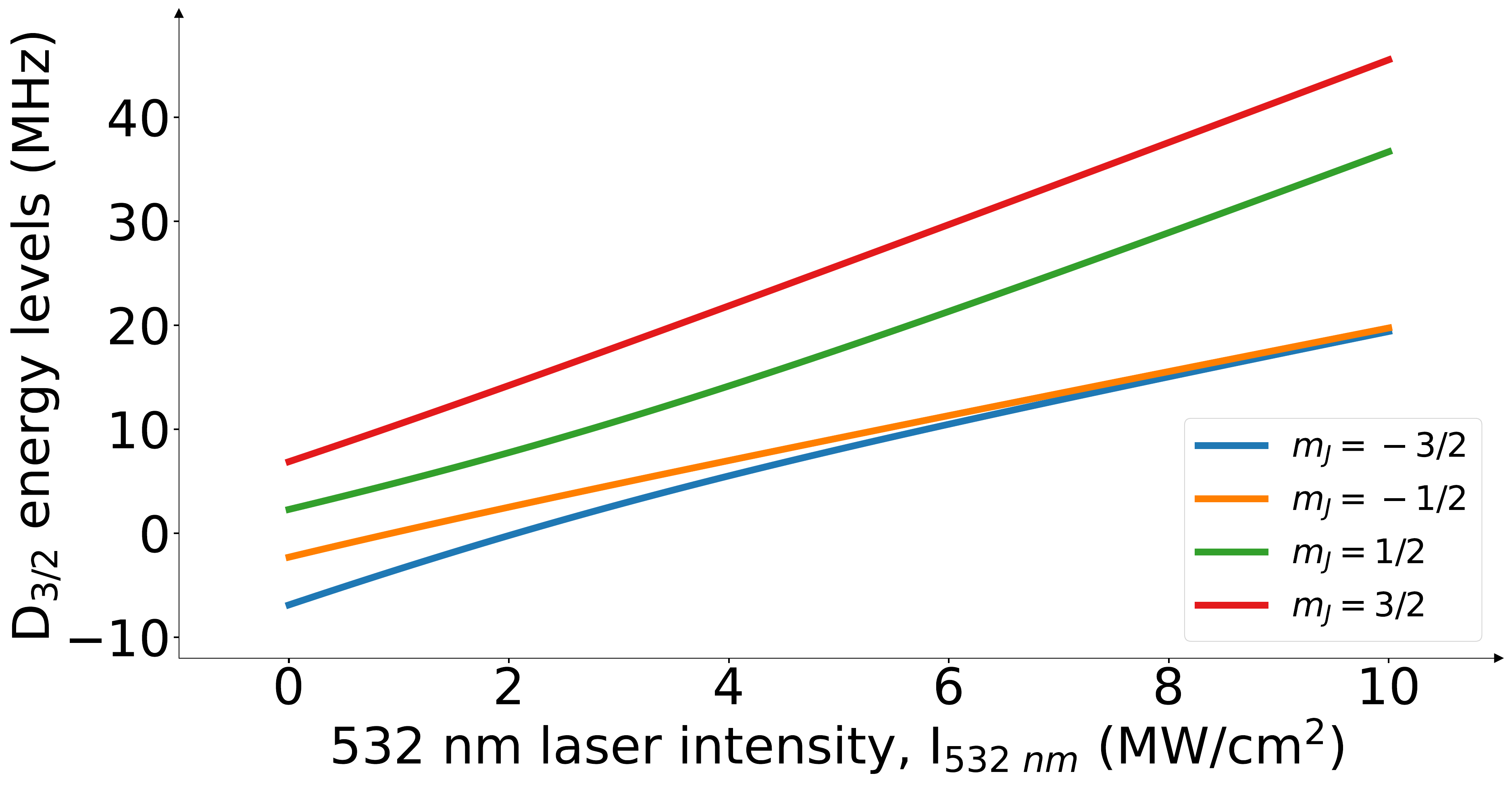}
\caption{\label{fig:Dshift_Ba138} AC Stark shift of \Ba. The external B field used here is 4.1~Gauss, which creates 4.59~MHz Zeeman splitting when there's no Stark shift. The zero energy is defined when both shifts are absent.}
\end{figure}

For atomic levels with $J=1/2$ (e.g. ground states $S_{1/2}$ and first excited states $P_{1/2}$ of Ba$^{+}$), the tensor polarizability vanishes and only the scalar shift exists, such that $|J, m_{J}\rangle$ (or $|F, m_{F}\rangle$ when non-zero nuclear spin) remain good quantum numbers. For levels with $J>1/2$ (such as $D_{3/2}$ and $D_{5/2}$ of Ba$^{+}$), the tensor part also exists and a new atomic basis needs to be computed before determining the shift because the operator does not commute with the Zeeman interaction. As an example, we diagonalize the overall Hamiltonian ($H=H_{AC\ Stark}+H_{Zeeman}$) of \Ba, and find the energy shift at different 532~nm laser intensity shown in Fig.~\ref{fig:Dshift_Ba138}.

\renewcommand{\arraystretch}{1.2}
\begin{table}[b]
\caption{\label{Table: SDP_Polarizability}%
532~nm polarizability of Ba$^{+}$. Transition wavelength and relevant $A_{J^{\prime}J}$ are from \cite{NIST_ASD}.
} 
\begin{ruledtabular}
\begin{tabular}{cccc}
\textrm{Level}&
\textrm{Considered transitions}&
\textrm{$\alpha_{nJ}^{scalar}$(a.u.)}&
\textrm{$\alpha_{nJ}^{tensor}$(a.u.)}\\
\colrule
$6S_{1/2}$ & 493, 455~nm & 549.75 & 0\\
$6P_{1/2}$ & \makecell{493, 650, 452, 389, 265, 253,\\ 220, 215, 201, 199, 189~nm} & 25.39\footnote{This value deviate a lot from \cite{UDportal} since leading contributions from first four considered transitions (almost) cancel out.} & 0\\
$5D_{3/2}$ & 650, 585~nm & -64.73 & 21.27\\
$5D_{5/2}$ & 614~nm & -84.81 & 84.81
\end{tabular}
\end{ruledtabular}
\end{table}

For atoms with non-zero nuclear spin like $^{137}$Ba$^{+}$ and $^{133}$Ba$^{+}$, there's one more term, i.e. hyperfine interaction, that needs to be taken into account. The overall Hamiltonian is:

\begin{equation}
\begin{aligned}
    H &=H_{AC\ Stark}+H_{Hyperfine} +H_{Zeeman} \\
      &=-\frac{1}{4}E^{2}[\alpha^{scalar}_{nJ} + \alpha^{tensor}_{nJ} \frac{6\boldsymbol{J_{x}}^{2}-2\boldsymbol{J}^{2}}{2J(2J-1)}] \\ & +\hbar A_{hfs}\boldsymbol{I\cdot J}+\hbar B_{hfs}\frac{6\boldsymbol{(I\cdot J)^{2}}+3\boldsymbol{I\cdot J}-2\boldsymbol{I^{2}}\boldsymbol{J^{2}}}{2I(2I-1)2J(2J-1)} \\& +\frac{g_{J}\mu_{B}}{\hbar} \boldsymbol{J\cdot B} \label{Overall Hamiltonian}
\end{aligned}
\end{equation}
where $A_{hfs}, B_{hfs}$ are hyperfine constants, and $g_{J}, \mu_{B}$ are $g-$factor and Bohr magneton respectively.

\begin{figure*}
\includegraphics[width=\textwidth]{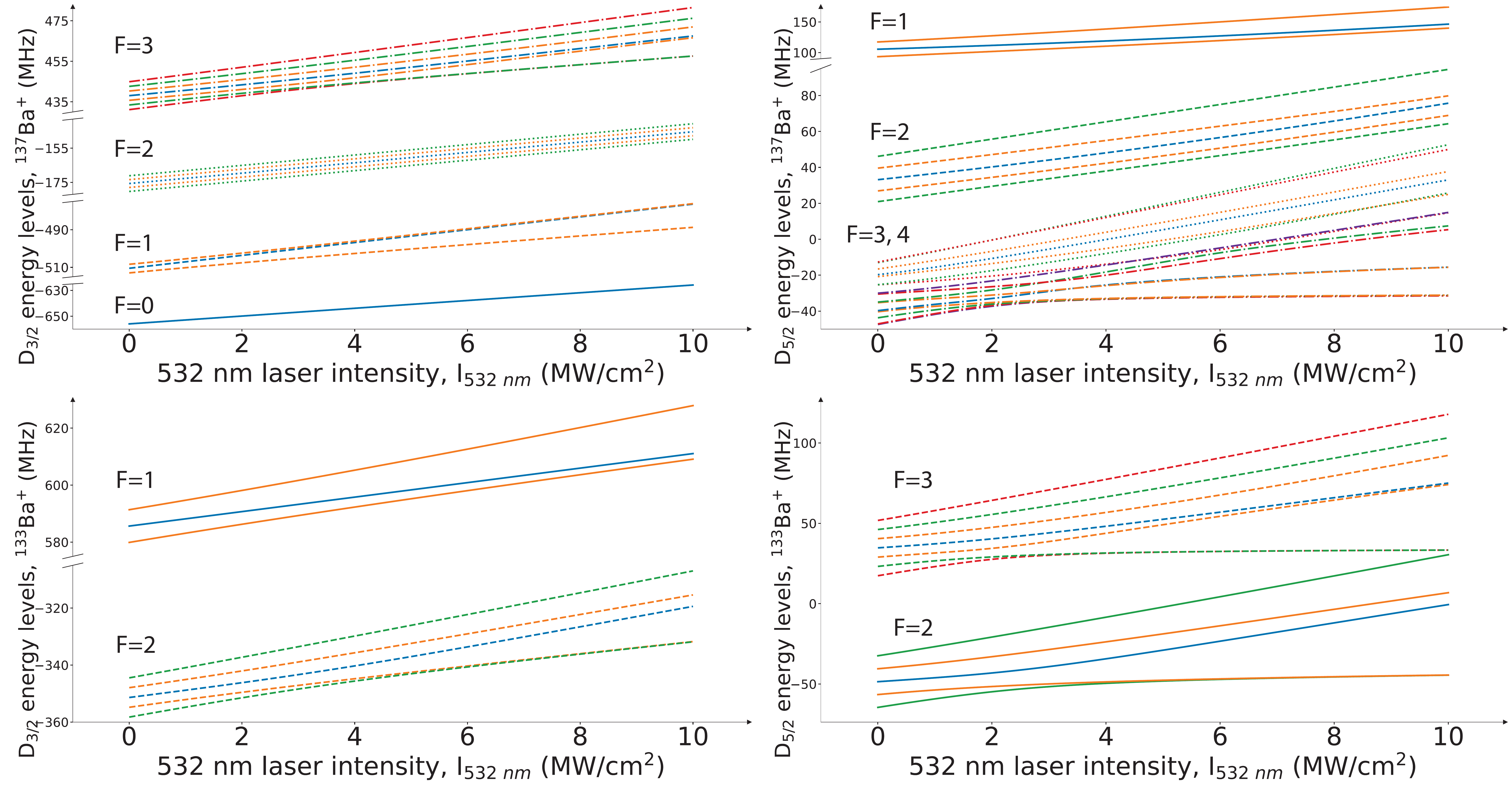}
\caption{\label{fig:133Ba137BaShift}
532~nm AC Stark shift of $^{137}$Ba$^{+}$ and $^{133}$Ba$^{+}$. Note that for $D_{5/2}$ of $^{137}$Ba$^{+}$, different Zeeman states ($F=3, 4$) are not fully resolvable even when AC Stark shift is absent. Here we assume an applied magnetic field of 4.1 Gauss which is used in the experiment and an intermediate B field regime where the Zeeman effect starts to interfere with hyperfine interaction\cite{low2025control}. As 532~nm intensity increases, $D_{5/2}$ sublevels will overlap and cross. Hyperfine constants used in this plot are from \cite{lewty2013experimental, hucul2017spectroscopy, christensen2020high} and zero energy are defined as the barycenter of the metastable $D$ level when there's no light.}
\end{figure*}

The $D_{3/2}$ and $D_{5/2}$ energy shift of $^{137}$Ba$^{+}$ and $^{133}$Ba$^{+}$ are shown in Fig. \ref{fig:133Ba137BaShift}. In the experimentally practical regime we consider in this paper (tens of mW to a few hundreds mW 532~nm laser power), the tensor AC Stark shift is comparable to Zeeman splitting ($\sim$ MHz for a few Gauss) but much smaller than $D_{3/2}$ hyperfine splitting of $^{137}$Ba$^{+}$/$^{133}$Ba$^{+}$ (hundreds of MHz), so state mixing will mostly happen among different Zeeman states within the same hyperfine level. However, states on $D_{5/2}$ levels of $^{137}$Ba$^{+}$/$^{133}$Ba$^{+}$ are different. Their relatively small hyperfine splitting indicate $D_{5/2}$ might be suitable only for electron-shelving with a small 532~nm AC Stark shift but not for driving cycling transitions for measurement with a large shift.

\section{\label{app:simulation} Numerical simulation of $S-D-P$ transitions}

In this section, we provide simulation results for $S-D-P$ transitions of auxiliary Ba$^{+}$ when 532~nm AC Stark shift is present. Although some states on $D_{3/2}$ are shifted closer to each other as shown in Fig.\ref{fig:Dshift_Ba138}, we show that by choosing 650~nm polarization and optimizing it's power and frequency, sufficient excitation on $P_{1/2}$ (0.04) could be achieved.

For \Ba, the time-dependent Hamiltonian in the rotating frame after rotating wave approximation (RWA) is:

\begin{equation}
\begin{aligned}
    H/\hbar &= \sum_{i=1}^{2}\omega_{S_{i}}|S_{i}\rangle\langle S_{i}| + \sum_{i=1}^{4}\omega_{\tilde{D}_{i}}|\tilde{D}_{i}\rangle\langle \tilde{D}_{i}| + \sum_{i=1}^{2}\omega_{P_{i}}|P_{i}\rangle\langle P_{i}|  \\
    &+ \frac{1}{2}\sum_{i, j}\Omega_{2052}^{(ij)}(1+e^{-i\delta t})|S_{i}\rangle\langle \tilde{D}_{j}| \\ &+ \frac{1}{2}\sum_{i, j}\Omega_{650}^{(ij)}|\tilde{D}_{i}\rangle\langle P_{j}|+h.c.
\end{aligned}
\end{equation}
where $\omega_{S_{i}}$, $\omega_{\tilde{D}_{i}}$, $\omega_{P_{i}}$ are energy of atomic states in the rotating frame, and $\Omega^{(ij)}$ are the Rabi frequencies. Here, we set a second 2052~nm tone whose frequency is detuned from carrier by $\delta$ because Rabi frequency $\Omega^{(ij)}_{2052}$ is usually much smaller than ground state energy splitting and addressing all the ground states is necessary for continuous fluorescence. Tilde represents new atomic eigenstates obtained from diagonalizing $H_{AC\ Stark}+H_{Zeeman}$.

The Rabi frequencies are determined by laser beam intensity and coupling strength between state $|i\rangle$ and $|j\rangle$. Because of state mixing, $|\tilde{D}_{i}\rangle$ states on $D_{3/2}$ are represented in the original basis ($|\tilde{D}_{i}\rangle=\sum_{m_{D}}c_{m_{D_{}}}^{i}|D_{3/2}, m_{D}\rangle$) before applying Wigner-Eckart theorem. We have:

\begin{equation}
    \Omega_{2052}^{(ij)} = \Omega_{2052}^{0}\sum_{m_{D}}c_{m_{D}}^{j}\sum_{q=-2}^{2}\begin{pmatrix}
1/2 & 2 & 3/2 \\
-m_{S_{i}} & q & m_{D}
\end{pmatrix}c_{kl}^{(q)}\epsilon_{k}\eta_{l}
\end{equation}

and 

\begin{equation}
    \Omega_{650}^{(ij)} = \Omega_{650}^{0}\sum_{m_{D}}(c_{m_{D}}^{i})^{*}\sum_{q=-1}^{1}\begin{pmatrix}
3/2 & 1 & 1/2 \\
-m_{D} & q & m_{P_{j}}
\end{pmatrix}c_{k}^{(q)}\epsilon_{k}
\end{equation}
where $c_{k}^{(q)}$, $c_{kl}^{(q)}$ are the first-order and second-order tensor, respectively\cite{james1998quantum}; $\eta_{l}$ is the unit wavevector, $\epsilon_{k}$ is the unit vector for laser polarization, and $\Omega_{650}^{0}$, $\Omega_{2052}^{0}$ are the base Rabi frequency which are the same for all transitions between two fine structure levels. Round matrices represent Wigner-3j symbols. Throughout all our simulations, we use $\epsilon_{k}\perp\eta_{l}\perp\vec{B}$. 

The collapse operators for 650~nm spontaneous emission are treated similarly:

\begin{equation}
    \hat{L}_{ij} =  \sqrt{2\Gamma_{650}}\sqrt{\sum_{m_{D}}(c_{m_{D}}^{i})^{*}\sum_{q=-1}^{1}\begin{pmatrix}
3/2 & 1 & 1/2 \\
-m_{D} & q & m_{P_{j}}
\end{pmatrix}^{2}} |\tilde{D}_{i}\rangle\langle P_{j}|
\end{equation}
where $\Gamma_{650}$ is the spontaneous emission rate, and $\hat{L}_{ij}$ for 650~nm photons of same polarization are added together.

\begin{figure*}
\includegraphics[width=\textwidth, height=6cm]{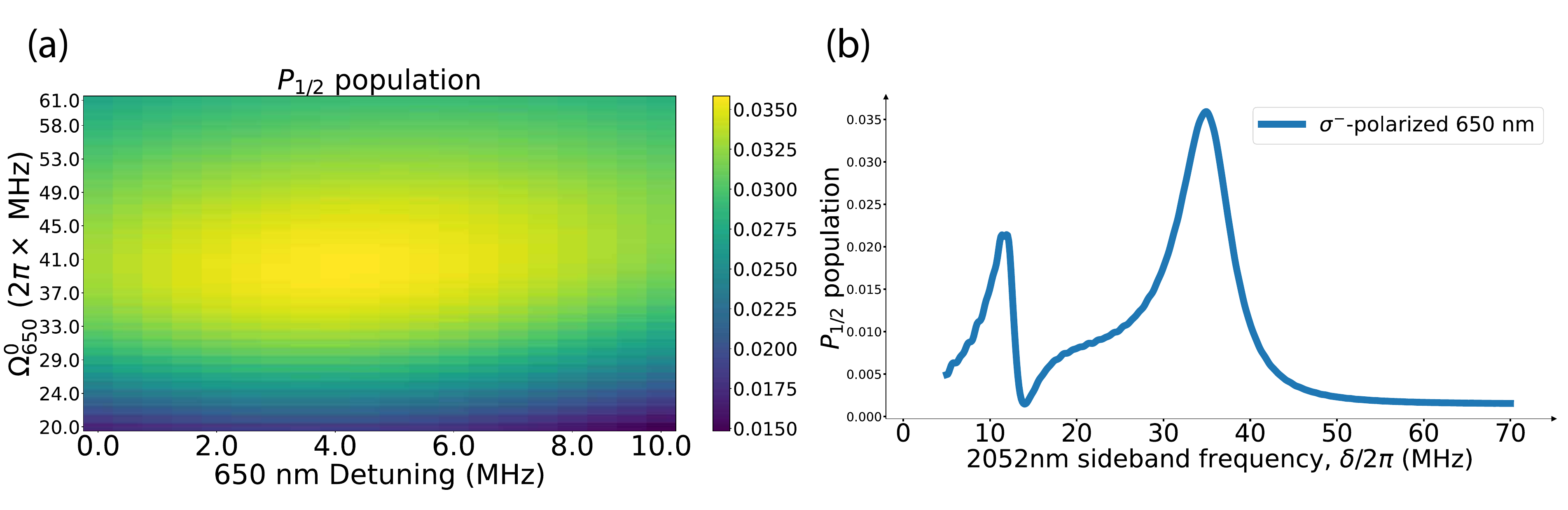}
\caption{\label{fig:Simulation138} \Ba\ numerical simulations. The 532~nm intensity used here is 6.1 MW/cm$^{2}$, which corresponds to more than 150~MHz AC Stark shift of $S_{1/2}$, and we choose $\Omega_{2052}^{0}$ such that the maximal value among all $\Omega_{2052}^{(ij)}$ is $2\pi\times 2$MHz. (a) 2D scan of 650 parameters. The 650~nm detuning is defined relative to the center frequency between $D_{3/2}$ and $P_{1/2}$. (b) $P_{1/2}$ population as a function of 2052~nm sideband frequency with optimal 650 parameters from (a). We set the carrier to drive the transition from $|S_{1/2}, m_{s}=1/2\rangle$ to $|\tilde{D}_{3/2}, m_{j}=-3/2\rangle$, and two peaks corresponds to the case where the sideband is resonant with transitions from $|S_{1/2}, m_{s}=-1/2\rangle$. The populations are calculated by taking an average after the dynamics become steady.}
\end{figure*}

The simulation results are shown in Fig. \ref{fig:Simulation138}. We implemented a 2D scan on 650 Rabi frequency and detuning ($\Omega_{650}^{0}$ and $\omega_{P_{i}}$) to find optimal 650 parameters. For each set of parameters, we also scan the frequency of 2052~nm sideband to maximize $P_{1/2}$ population. Even circularly polarized 650~nm used here can repump $D_{3/2}$ when tensor shifts are comparable to Zeeman splitting, and closer atomic states are not preventing efficient repumping.

\begin{figure}
\includegraphics[width=0.48\textwidth]{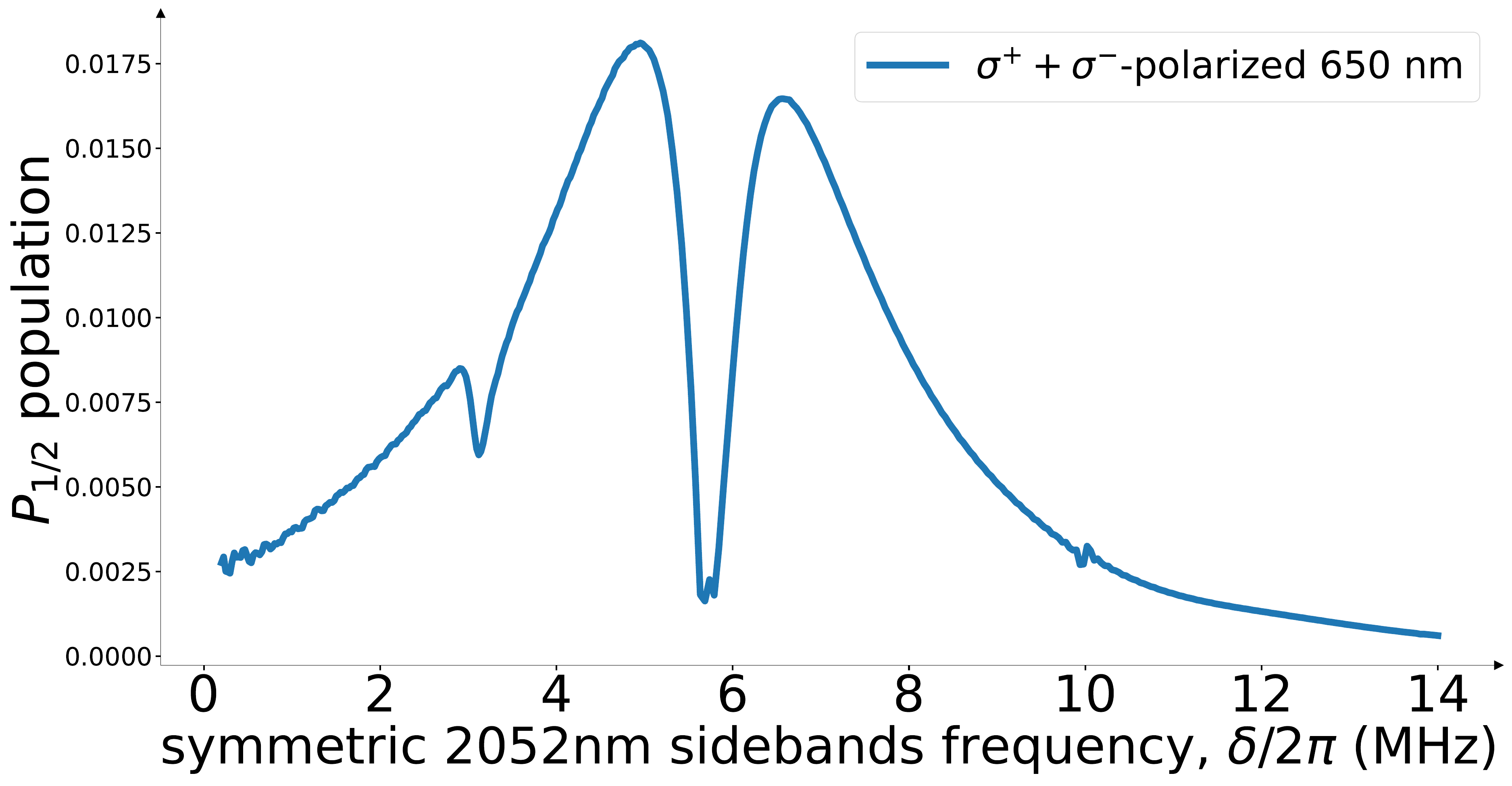}
\caption{\label{fig:Simulation137} $^{137}$Ba$^{+}$ numerical simulations. 532~nm intensity and $\Omega_{2052}^{0}$ are chosen in a same way as in Fig. \ref{fig:Simulation138}. The dark resonance around 6~MHz happens because two sidebands are driving two ground states to the same metastable state on $\tilde{D}_{3/2}$. It is avoided by changing 2052~nm frequencies.}
\end{figure}

To further illustrate the feasibility of our MCMR scheme on ions with non-zero nuclear spin, we implemented simulation on $^{137}$Ba$^{+}$ as shown in Fig. \ref{fig:Simulation137}. Its complicated energy levels require more 2052~nm tones to address all states on ground level; and require multiple 650~nm tones to address different $D_{3/2}$ hyperfine levels. For simplicity and simulation efficiency, we assume that there is a 2052~nm carrier and two symmetric sidebands (detuned by $\delta$ from the carrier) that drive transitions from $|S_{1/2}, F=1\rangle$ to $|\tilde{D}_{3/2}, F=2\rangle$; and two additional symmetric sidebands, which are $2\delta$ apart from each other, that drive transitions from $|S_{1/2}, F=2\rangle$ to $|\tilde{D}_{3/2}, F=2\rangle$. Note that this is not necessarily optimal for maximal $P_{1/2}$ excitation, and the result here (0.02) can be improved by optimizing 2052~nm sidebands and 650~nm parameters (detuning, polarizations, and Rabi frequencies). We leave it for future work.

\section{\label{MCMR time} $T_{MCMR}$ estimation}

In this section we explain how the mid-circuit measurement and reset time, $T_{MCMR}$, is estimated.

The measurement time of the auxiliary ion is inversely proportional to photon detection rate $R_{detection}$, which is determined by the collection efficiency ($\varepsilon$), spontaneous emission rate ($\Gamma$), and steady state population of the excited state ($f$): 

\begin{equation}
\begin{aligned}
    R_{detection} & = \varepsilon\cdot f\cdot\beta\Gamma \\
      &=\varepsilon_{lens}\varepsilon_{fiber}\varepsilon_{detector}\cdot f\cdot\beta\Gamma
\end{aligned}
\end{equation}
where $\beta$ is the branching ratio, and $\varepsilon_{lens}$, $\varepsilon_{fiber}$, $\varepsilon_{detector}$ are collection percentage of the imaging lens, fiber coupling efficiency, and detector quantum efficiency, respectively.

The state-of-the-art result (11~$\mu$s) in \cite{crain2019high} was using a single 0.6~NA lens and a superconducting nanowire single photon detector (SNSPD) to collect 370~nm photons from $^{171}$Yb$^{+}$. Their steady state $P_{1/2}$ population is 0.1, which is calculated from provided 370~nm laser intensity and the formula in their Methods. By assuming the same collection efficiency $\varepsilon$ and using the spontaneous emission rate of Ba$^{+}$ instead of Yb$^{+}$ ($\Gamma=2\pi\times20.5$MHz, $\beta=0.732$ for Ba$^{+}$, while $\Gamma=2\pi\times19.6$MHz, $\beta=0.995$ for Yb$^{+}$), the $R_{detection}^{Ba}$ would be 30\% of $R_{detection}^{Yb}$. The $P_{1/2}$ population we used here is $f=0.04$ for \Ba. This lower photon detection rate leads to a longer detection time of 36~$\mu$s. In practice, the measurement time could be shorter since fibers and detectors for collecting 493~nm photons ($\varepsilon_{fiber}, \varepsilon_{detector}$) usually have higher efficiency.

\section{\label{532stability}532 setup and stability}
We now describe the setup and likely sources of instability of our 532~nm Stark-shifting beam. For the results presented here, we combined both the detection and the Stark-shifting beam in a cage-mounted setup attached to the top of the ion trap chamber. 
The 532~nm laser light transmits vertically downwards, perpendicular to the magnetic field, and traverses a 532/493~nm dichroic and a NA = 0.6 objective for tight focusing. We note this beam is elliptical, with the longer part aligned axially along the trap, due to mis-alignment of the imaging objective. 
The 532~nm light is not power stabilized, and we use motorized control of the stage to align the beam to the ion to $<1~\mu$m precision.
The 493~nm ion florescence transmitting upwards reflects off the dichroic and is collected into a $50~\mu$m multimode fiber and detected by an avalanche photodiode (APD).

To investigate the stability of this setup, we made an equal superposition of the ground state qubit using 532~nm Raman and repeatedly measured the population at low power ($\sim 3$~mW) over 18 minutes. We measured a slow, periodic drift with a period of about 6 minutes. This contributes to calibration error in our experiments. Next, we checked the stability of the polarization of the 532~nm light, which we found to be stable to $<1$~dBm. 
We also investigated the power coming out of the fiber at the chamber, but found power fluctuations $< 3 \%$. 
To measure fast power fluctuations, we put the light from the fiber onto a fast photodiode and took the fast Fourier transform. We found low frequency peaks on the order of 400~kHz that could explain some of the instability we see in our experiments.

\begin{figure}
\includegraphics[width=0.48\textwidth]{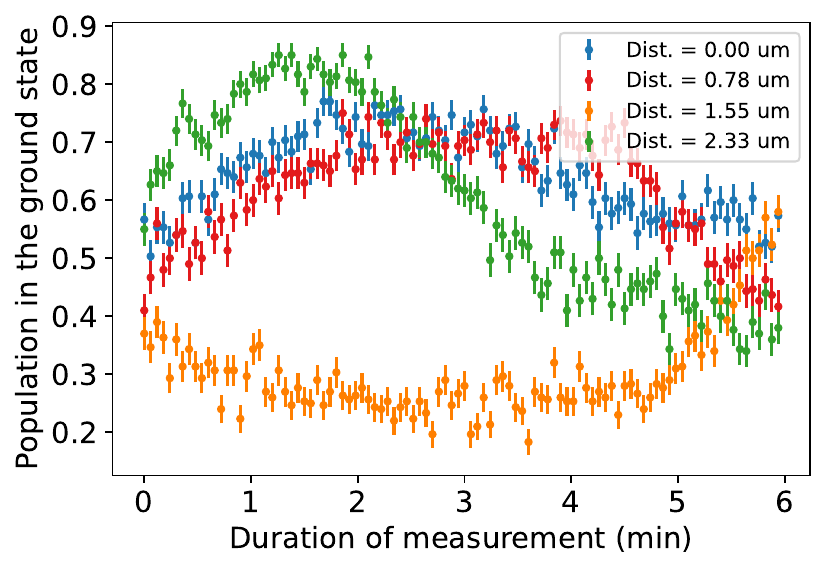}
\caption{\label{fig:532Stabilitypop}
Measuring the ground state population in an initial superposition over 6 minutes at different distances from center alignment to the ion. By repeatedly measuring the population over time, we can see the drift in population caused by 532~nm instability.
The variation in population over this timescale is largely the same until the edge of the beam, which should be more sensitive to pointing noise. We convert this variation in population to Rabi rate variation, from which we can calculate the resultant deviation in AC Stark shift.
All error bars are derived using the standard deviation of a binomial distribution, where each point has 300 shots.}
\end{figure}

In addition, we repeatedly measured the population over 6 minutes, when in an initial superposition, with the 532 nm beam at different distances from the center of the ion. 
The results of these measurements is shown in Fig.~\ref{fig:532Stabilitypop}.
We observe both a slow drift (3-6 min) and fast fluctuations ($5-10$~ms) in the measured population.
There was no noticeable difference in the population fluctuations or the period of the slow drift up to $1.55\mu$m from center. 
However, near the edge of the beam, we see a larger fluctuations and the period of drift halved. 
We converted the difference in population to a variation in AC Stark shift.
The resulting standard deviation in AC Stark shift for distances between $0-1.55\mu$m is about $12 \%$ of the total shift applied, but at $2.33\mu$m it is about $17 \%$.

We model the Stark shift deviation as an effective laser linewidth that broadens both 2052~nm and 650~nm transitions, and run the simplified $\{S_{1/2}, D_{3/2}, P_{1/2}\}$ simulation. We find that the bright state fidelity is 96.8\% - 94.6\% when the AC Stark shift is 1.5~MHz; and is 90.1\% - 83.1\% when the AC Stark shift is 3.6~MHz. The higher and lower fidelity corresponds to 12\% and 17\% deviation, respectively. These simulated numbers agree with our experimental results (97.0(5)\% for 1.5~MHz shift and 87.5(10)\% for 3.6~MHz shift) and indicate that unstable beam pointing is the main reason that leads to decreasing fidelity.

Due to these Stark shift fluctuations, we could only apply a Stark shift up to 5~MHz before the noise was too high to resolve spectroscopic features. Thus, we chose a smaller Stark shift to test our MCMR protocol. 

\section{\label{app:2052 set up} 2052 nm setup}
We empirically measure the 2052 nm 90 kHz linewidth over a few seconds by performing spectroscopy and looking at the FWHM. Note that this is not the instantaneous linewidth.
We use a low-finesse cavity $F<1000$ from Thorlabs to lock 2052~nm. Future implementations would utilize a high-finesse cavity and a better 2052 nm laser with more power and stability.

When we first applied both tones of 2052 nm, we simply applied the two-tone laser field simultaneously and continuously. However, driving the AOM in this way was very inefficient and made the 2052~nm power very low. We then tried switching between each tone every 5 us to get more power, but the wings in the Gaussian profile caused off-resonant driving to other transitions. Thus, we implemented Blackman pulse shaping while switching between the two tones because Blackman pulse shaping eliminates the sidelobes that are present in Gaussian beams.

Additionally, the 2052~nm beam was optimized in both geometry and polarization such that it couples most efficiently for $\Delta m=0$ transitions. 

\bibliographystyle{apsrev4-2}
\bibliography{MAIN}

\end{document}